# Deep Learning Model for Demodulation Reference Signal based Channel Estimation

Yu Tian*, Chengguang Li*, Sen Yang

*Abstract*—In this paper, we propose a deep learning model for Demodulation Reference Signal (DMRS) based channel estimation task. Specifically, a novel Denoise, Linear interpolation and Refine (DLR) pipeline is proposed to mitigate the noise propagation problem during channel information interpolation and to restore the nonlinear variation of wireless channel over time. At the same time, the Small-norm Sample Cost-sensitive (SSC) learning method is proposed to equalize the qualities of channel estimation under different kinds of wireless environments and improve the channel estimation reliability. The effectiveness of the propose DLR-SSC model is verified on WAIC Dataset. Compared with the well know ChannelNet channel estimation model, our DLR-SSC model reduced normalized mean square error (NMSE) by 27.2dB, 22.4dB and 16.8dB respectively at 0dB, 10dB, and 20dB SNR. The proposed model has won the second place in the 2nd Wireless Communication Artificial Intelligence Competition (WAIC). The code is about to open source.

*Index Terms*—wireless communication, deep learning, channel estimation

## 1 Introduction

Accurate channel estimation is essential to ensure the performance of wireless communication systems. The channel information in wireless communication systems can be expressed as a matrix, in which each value represents the channel response at a given time slot and frequency bin. One of the core issues of channel estimation is Demodulation Reference Signal (DMRS) based channel estimation [1]. DMRS occupies a small amount of positions on the time-frequency plane and is sent with data signal at the transmitter. The time-frequency positions and values of DMRS are known to receiver. At the receiver, by comparing the transmitted and received DMRS, the DMRS sub channel matrix can be obtained. The objective of DMRS channel estimation is to derive channel responses at all time-frequency positions based on DMRS sub channel matrix and restore the full channel matrix. Currently, DMRS channel estimation is mainly realized by linear algorithms such as Wiener filtering. From the perspective of the actual system, this type of linear algorithm is effective, but it is not the optimal algorithm. Introducing Artificial Intelligence (AI) into DMRS channel estimation can further improve the accuracy of channel estimation, thereby improving the performance of the communication system.

Recently, Deep Learning (DL) method has been reported for DMRS channel estimation [5]. In [5], the DMRS channel matrix is modeled as a 2D low-resolution (LR) image, and a pipeline consist of super resolution (SR) and image restoration (IR) is proposed to restore the full channel matrix. However, this SR-IR pipeline may suffer from noise propagation problem. Since DMRS channel matrix is noise-corrupted, the SR operation may spread noise to other time-frequency positions, which will make subsequent IR operation very challenging.

It should be noted that different channel environments have different tolerance to channel estimation errors. For wireless channels with severe shadow effects, values in the channel matrix are small, a small absolute error may significantly change the characteristics of channel effects. Thus, high estimation accuracy is required. In contrast, wireless channels that are less affected by the shadow effect do not require such high estimation accuracy.

In this paper, we present a Denoise, Linear interpolation and Refine (DLR) pipeline and (SSC) learning method for high quality DMRS channel estimation. In DLR pipeline, the DMRS channel matrix is firstly denoised to prevent noise spreading. Then the denoised DMRS channel matrix is linearly interpolated to coarsely estimate the full channel matrix. At last, the coarsely estimated full matrix is refined to restore nonlinear relationship between different time-frequency positions. The SSC learning aims to achieve differentiated processing and guide channel estimation model to allocate more model capacity to fit wireless channels with shadow effects. As a result, the estimation qualities for different wireless channel environment can be equalized and the reliability of channel estimation can be effectively improved.

## 2 Principle

### 2.1 DLR Pipeline

The pipeline of proposed DLR wireless channel estimation model is shown in Fig 1. The sub channel matrix at DMRS positions $\tilde{\mathbf{H}}_{\text{DMRS}}$ is firstly fed into Denoise module, which aims to suppress noise in $\tilde{\mathbf{H}}_{\text{DMRS}}$, and prevent noise from spreading to other time-frequency positions. The denoised DMRS channel matrix is represented as $\mathring{\mathbf{H}}_{\text{DMRS}}$. Assume that the channel information at different time-frequency positions is quasi linearly correlated. On this basis, the sub channel matrix at non-DMRS positions can be represented as follows.

*These authors contributed to the work equally and should be regarded as co-first authors. Yu Tian is with the Qian Xuesen Laboratory of Space Technology, Beijing, China (tianyu@qxslab.cn). Chengguang Li is with the School of Aerospace Science and Technology, Xidian university, Xi'an, China (lichengguang@stu.xidian.edu.cn). Sen Yang is with the School of Information Science and Technology, Peking University, Beijing, China (magicys@qq.com)

$$\mathring{\mathbf{H}}_{\text{non-DMRS}} = \mathring{\mathbf{H}}_{\text{non-DMRS}}^{\text{linear}} + \mathring{\mathbf{H}}_{\text{non-DMRS}}^{\text{nonlinear}}$$

The first term is linearly correlated with $\mathring{\mathbf{H}}_{\text{DMRS}}$, and the second term is nonlinearly correlated with $\mathring{\mathbf{H}}_{\text{DMRS}}$. In Linearization module, $\mathring{\mathbf{H}}_{\text{non-DMRS}}^{\text{linear}}$ is derived from $\mathring{\mathbf{H}}_{\text{DMRS}}$ by means of linear interpolation, then $\mathring{\mathbf{H}}_{\text{non-DMRS}}^{\text{linear}}$ and $\mathring{\mathbf{H}}_{\text{DMRS}}$ are combined to get the full channel matrix $\mathring{\mathbf{H}}$. At last, the Refine module is used to refine $\mathring{\mathbf{H}}$ and add nonlinearity for non-DMRS positions. Additionally, normalization and denormalization is added to accelerate model convergence.

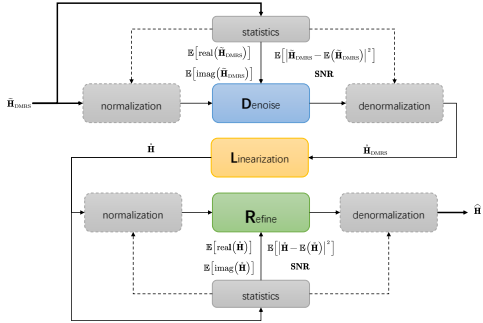

Fig 1. DLR Pipeline for DMRS channel estimation

The backbone of Denoise module is based on Uformer [2]. Compared with vanilla Uformer, our Denoise module effectively reduce the computational requirement and speedup training. Assume an input with shape of $C \times W \times H$, where $C$, $W$, and $H$ represent feature number, spatial width and spatial height. Uformer applies window partition and Window-based Self Attention (WSA) on 2D spatial plane. Each window contains $HW$ tokens. However, the Denoise module treats spatial height as additional features, and reshape the input as $CH \times W$. Then 1D-WSA is applied, which only results in $W$ tokens in each window. Given $N$ tokens, WSA requires to calculate pair-wise correlation between tokens, which requires $O(N^2)$ complexity. With the token number reduced by $H$, the computation complexity of Denoise module can be reduced by a factor of $H^2$.

The Architecture of Denoise module is shown in Fig 2, which is composed of four stages. In each stage, two DN blocks are employed. DN block consists of a WSA block and a Depthwise Separable Convolutions (DSC) block. Given an input with shape of $dim \times W$, WSA divides the input as several non-overlapping window with shape of $dim \times win$. In each window, Multi Head Self Attention (MHSA) block is applied over $win$ tokens. Then a Squeeze and Excitation (SE) block is applied to perform dynamic feature recalibration. At last, all windows are combined to resume the input shape. Between different stages, a Down & Expand block or Up & Squeeze block is inserted. The function of Down & Expand block is to down sample the frequency dimension and expand channel depth. The down sample and expand factor are set to 2. The Down & Expand block is implemented as by a 1D convolution operation with kernel size of 4 and stride of 2. The function of Up & Squeeze block is to up sample the frequency dimension and squeeze channel depth. The up sample and squeeze factor are set to 2. The Up & Squeeze block is implemented as by a 1D transposed convolution operation with kernel size of 2 and stride of 2.

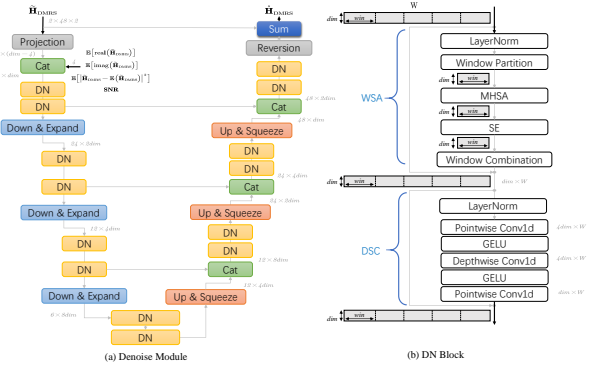

Fig 2. Architecture of Denoise module

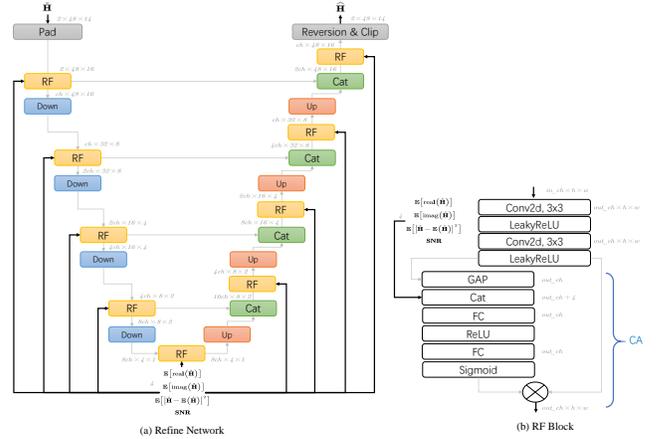

Fig 3. Architecture of Refine module

The Refine module adopts an Unet [3] architecture to exploit the time-frequency relationship in channel matrix, in order to restore the nonlinear changes in time dimension. The architecture of Refine module is shown in Fig 3. In encoder branch, Downsample block and RF block are alternatively used in order to expand the receptive field and extract features at different scales. At the bottleneck layer, global context information over all time-frequency positions is be captured, which implicitly encodes the information such as user moving speed and Doppler effect. The global context information can be used to infer the time coherence of wireless channel. Afterwards, the decoder branch alternatively adopts Upsample block and RF block to upsample the time-frequency plane and restore details at higher resolution. The Downsample block is implemented by max pooling. The Upsample block is implemented by bilinear interpolation. The downsample factor and upsample factor are both set to 2.

## 2.2 SSC Learning

In order to comprehensively consider the different requirements of different channel environments for channel estimation, the normalized mean square error (NMSE) is typically used to measure the quality of channel estimation. The calculation method of NMSE is as follows.

$$NMSE(\mathbf{H},\hat{\mathbf{H}}) = \frac{\|\mathbf{H}-\hat{\mathbf{H}}\|_2}{\|\mathbf{H}\|_2} = \frac{MSE(\mathbf{H},\hat{\mathbf{H}})}{\|\mathbf{H}\|_2}$$

where $\mathbf{H}$ and $\hat{\mathbf{H}}$ respectively represents the ground truth and estimation of channel matrix. $\|\ \|_2$ denotes the norm of matrix and can be used to indicate the strength of the shadow effect.

As can be seen, depending on channel matrix norm, estimation MSEs contribute differently to NMSE. For channel matrices with small norms, the estimation MSEs will be amplified, and produce great negative impact on NMSE score.

In order to deal with above problem, we propose the Small-norm Sample Cost-sensitive (SSC) learning method. The SSC learning aims to guide channel estimation model to achieve differentiated processing and equalized channel estimation qualities for channel matrices with different norms.

The SSC learning is implemented through Channel Statistics Information Fusion (CSIF) and Norm Aware Oversample Strategy (NAOS). CSIF is used to help channel estimation model to better recognize small-norm channel matrices by incorporating channel statistics. NAOS is used to guide the estimation model to learn more features for small-norm channel matrices in order to better estimate them.

CSIF is implemented by incorporating channel statistics into the Denoise module and Refine module in DRL pipeline. In Denoise module, the expectation and variance of channel matrix, and SNR are concatenated with the embedding of the input. In Refine module, the statistics of channel matrices are imported into the RF block via Channel Attention (CA) [4], which provides a mechanism to select different features for different kinds of channel matrices in a soft decision manner.

The NAOS uses matrix norm to change the sampling probability of different channel matrices, so that small norm channel matrices can appear with higher probabilities during the training process. This equivalently increases the weight of small channel matrices in loss, which enables the network to better fit small channel matrices and reduce their recovery MSE. The formula of sampling probability is follows.

$$P_i = \frac{bias - \log_{10}^{\|\mathbf{H}\|_2}}{\sum_i bias - \log_{10}^{\|\mathbf{H}\|_2}}$$

Lowering the bias can give higher sampling weight to small norm channel matrices.

## 3 Experiment

### 3.1 Setup

We verify the effectiveness of our DRL-SSC model on WAIC Dataset. The WAIC Dataset put forwards a DMRS channel estimation task on a time-frequency plane with 96 frequency bins and 14 time slots. It contains 210000 samples. Each sample consists of a DMRS sub channel matrix and the ground truth of corresponding full channel matrix. The ground truth is with shape of $96 \times 14 \times 2$, which correspond to 96 frequency bins, 14 time slots, the real part and the imaginary part. The DMRS sub channel matrix is with shape of $48 \times 2 \times 2$, which correspond to 48 frequency bins, 2 time slots, the real part and the imaginary part. The DMRS signal occupies the odd frequency bins, and the 4th and 12th time slots. Each sample is generated with random delay spread, UE speed and SNR. The SNR varies from 0 dB to 20 dB. For each SNR value, 10000 samples are generated.

The proposed DRL-SSC channel estimation model is implemented with Pytorch. In Denoise module, the embedding dimension of input $dim$ is set to 32. The window sizes $win$ are set to [8, 4, 2, 2] for four stages. The learning rate is set to 1e-4 with the batch size of 400. The channel factor $ch$ in Refine module is set to 32. The bias of NAOS is set to 5. The training epochs is set to 2400. We use Adam optimizer and L1 loss. In order to reduce the model size to fit in the edge computing device, we save the final model weights as half-precision floating-point format, which can reduce the model size by half. Dataset is divided into training set and validation set according to the ratio of 9:1. The model was trained on a NVIDIA V100 32GB GPU and an Intel(R) Xeon(R) Silver 4116 CPU @ 2.10GHz. ChannelNet [5] is chosen as the baseline model. As far as we know, this is the state-of-the-art deep learning model for DMRS channel estimation task. In order to facilitate comparison, we have expanded the capacity of ChannelNet so that it has roughly the same output as our model.

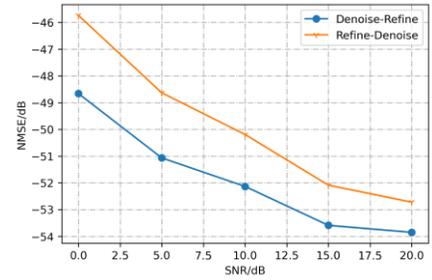

Fig 4. The effectiveness of DLR pipeline

### 3.2 Results

Firstly, we evaluate the effectiveness of DLR pipeline by swapping the positions of Denoise module and Refine module. In Refine-Denoise pipeline, the DMRS sub channel matrix is firstly linearly interpolated and refined by Refine module. As a result, the noise is spread all over the time-frequency plane. As can be seen in Fig 4, the Refine-Denoise pipeline severely degrades the estimation performance.

In order to verify the effectiveness of SSC learning, we respectively remove CSIF and NAOS. The results are compared in Fig 5. As can be seen, without CSIF, the performance of channel estimator degrades significantly. This indicates channel statistics information plays an important role the channel estimation model.

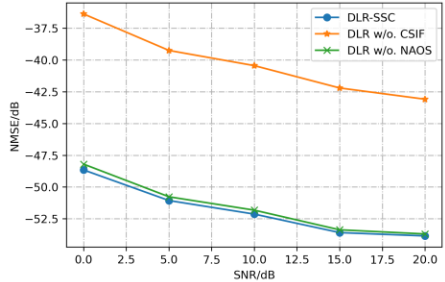

Fig 5. The effectiveness of SSC learning

The performance comparison between the ChannelNet and our DLR-SSC model is shown in Fig 6. Results shown that Our model significantly reduces the NMSE of channel estimation. At all SNRs levels, the estimation NMSEs of our model are all lower than -45dB. The advantage of our model becomes more obvious in the low signal-to-noise ratio area. Compared with ChannelNet, our model reduces the NMSE by 16.8dB and 27.2dB for 20 dB and 0dB SNR.

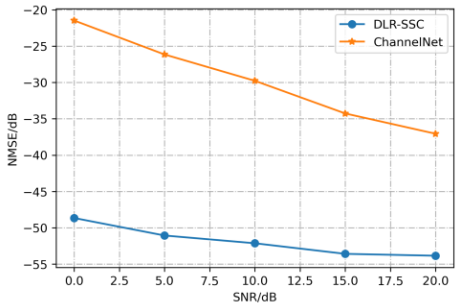

Fig 6. Comparison with ChannelNet

The prediction result and ground truth are shown in Fig 7. Each row represents a different sample. The first column represents the prediction results given by ChannelNet. The second column represents the prediction results given by DRL-SSC model. The last column represents the ground truth. By exploiting the time-frequency relationship in channel matrix, DRL-SSC model gives much smoother prediction results.

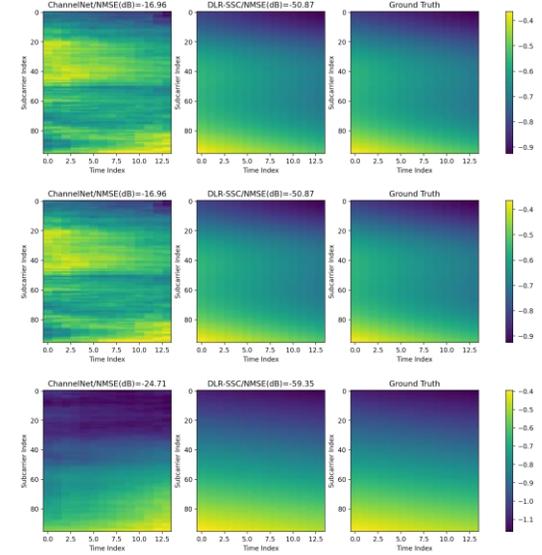

Fig 7. Prediction results

## 4 Conclusion

In this paper, we present the DLR-SSC model for high quality DMRS channel estimate. In DLR pipeline, we utilize the Uformer backbone to denoise DMRS channel matrix, and Unet architecture to exploit the time-frequency relationship in channel matrix, so as to restore the nonlinear changes in time dimension. Through theoretical analysis we found that small-norm channel matrices produce great negative impact on the quality of channel estimate. To solve this problem, we propose the Small-norm Sample Cost-sensitive (SSC) learning method. SSC learning optimizes the model training by incorporating channel statistics information fusion and norm aware oversample strategy, which will facilitate channel estimation model to better recognize and restore small-norm channel matrices. Experiment results show that the proposed method can effectively improve the quality and reliability of channel estimation under various wireless channel conditions.